\documentclass[epj]{svjour}
\usepackage{graphics}
\begin{document}
\title{Entrenched time delays versus accelerating opinion dynamics}
\subtitle{Are advanced democracies inherently unstable?}
\author{Claudius Gros} 
%
%
\institute{Institute for Theoretical Physics, Goethe University Frankfurt, 
Frankfurt am Main, Germany}

\date{September 2017}
%
\abstract{
Modern societies face the challenge that the time scale of opinion formation is
continuously accelerating in contrast to the time scale of political decision
making. With the latter remaining of the order of the election cycle we examine
here the case that the political state of a society is determined by the
continuously evolving values of the electorate. Given this assumption we show
that the time lags inherent in the election cycle will inevitable lead to
political instabilities for advanced democracies characterized both by an
accelerating pace of opinion dynamics and by high sensibilities (political
correctness) to deviations from mainstream values. Our result is based on the
observation that dynamical systems become generically unstable whenever time
delays become comparable to the time it takes to adapt to the steady state. The
time needed to recover from external shocks grows in addition dramatically
close to the transition. Our estimates for the order of magnitude of the
involved time scales indicate that socio-political instabilities may develop
once the aggregate time scale for the evolution of the political values of the
electorate falls below 7-15 months.  
%
%
} 
\maketitle
%
\section{Introduction}

A dynamical system with time delays reacts not
only to its current state, but also to what occurred
in the past. It is well known in this context that
time-delayed dynamical systems are prone to instabilities 
whenever the delay times become comparable to the time 
scales needed to react to current events and
perturbations \cite{erneux2009applied,gros2015complex}. 
To give an example from economy, consider just-in-time (JIT) 
manufacturing, for which the time scales regulating the 
delivery process are typically of the order of hours 
\cite{singh2012just}. Even small perturbations in the
supply chain would lead to an immediate break down 
of JIT manufacturing, as a whole, if the management of 
the involved companies would need days or weeks to react 
to an outage.

The dynamics of democratic political systems shares
certain basic similarities to manufacturing processes
like just-in-time manufacturing, with the political
institutions (parliament, government) reacting to
shifts in the demands of the electorate \cite{schnellenbach2015behavioral}.
It has been noticed in particular that the temporalities 
of economy and culture are driven by every faster cycles 
of innovation, change and replacement \cite{wolin1997time},
with political time remaining on the other side high
\cite{goetz2014question}.
There is hence an evolving mismatch of the speed 
of formal democracy \cite{fleischer2013time} with regard 
to the accelerating speed of capital \cite{tomba2014clash}, 
of economic decision making, opinion dynamics 
\cite{wolffsohn2001nomen} and of modern life in 
general \cite{rosa2013social}. 

The ongoing differentiation of societal time scales,
with opinion dynamics accelerating in contrast to
institutional decision making, did manifest itself 
in several political developments occurring
in 2016/2017. In French politics, to give an example,
electorate values changed so fast that the `En Marche'
movement could raise in essentially a single year from 
nowhere to a center role in French politics 
\cite{pain2017unusual}. The extended time scale of 
three or more years, as presently envisioned, to carry 
out the 2016 popular vote in favor of a Brexit 
\cite{menon2016brexit}, is on the other side exemplary 
for the prolonged time political institutions need 
to react to demands of the electorate. Our aim is 
here to develop a framework describing conflicts 
of temporalities on a basis that abstracts from 
specific circumstances. Our approach is particularly 
well suited for advanced democracies, i.e.\ for societies 
in an advanced state of acceleration.

Modern democracies are characterized additionally both
by an increasing level of skepticism towards political 
institutions \cite{dalton2004advanced} and by the ongoing
refinement of political correctness norms \cite{hughes2011political,maass2013does}.
This continuously increasing sensitivity to deviations from 
the mainstream normative order has its equivalent in economics, 
where companies not adhering to normative standards of 
reliability will if find difficult, in a world dominated by 
JIT manufacturing, to build up profitable business relations. 
Here we show that fine-tuned political correctness norms are 
directly related to the underlying acceleration of societal 
responses. Fast opinion dynamics and a high level of political 
correctness are in our model both indicative of political 
systems close to a dynamical instability. Fine-tuning a 
political system reduces consequently its robustness against 
perturbations.

\section{Model}

We denote with $D=D(t)$ and with $V=V(t)$ 
aggregate variables measuring the state of the democracy
and of the values of the electorate, its cultural
dimension \cite{abdollahian2012dynamics}. The time $t$ will
be measured in years. We remain here on a relative abstract 
level, noting however that standard country-specific indicators 
\cite{spaiser2014dynamics,alexander2012measuring} 
for both democracy and values may be taken as proxies for 
$D$ and $V$. Alternatively one may consider the level of
economic development, instead of the cultural dimension,
as the basic variable interacting with the state of
the democracy \cite{ranganathan2014bayesian}. 

A political system is democratic, per definition, whenever
$D(t)$ is reactive to changes in the values $V(t)$ of the 
electorate. This relation is captured by
\begin{equation}
T_D \frac{d}{dt} D(t) \ =\   V(t-T) - D(t)~,
\label{eq:dot_D}
\end{equation}
where $T_D$ denotes the time democratic institutions need to 
aligns themselves to the demands expressed by the electorate.
There is however an additional time scale involved, the time lag 
$T$. Time lags arise on one side from the circumstance that 
the electorate has to wait in a representative democracy on the 
average several years before it can express its value forcefully 
at election time \cite{goetz2014question}. Time lags also
occur generically in political decision making. It will take
about three years, if at all, to implement popular will in the 
case of the Brexit \cite{inglehart2016trump}.

The overall process modeled by (\ref{eq:dot_D}) describes a 
highly idealized democracy. We note, however, that the 
intricacies of real-life political decision making will 
enhance the effect here studied.

\begin{figure}[t]
\centering
\resizebox{0.9\columnwidth}{!}{\includegraphics{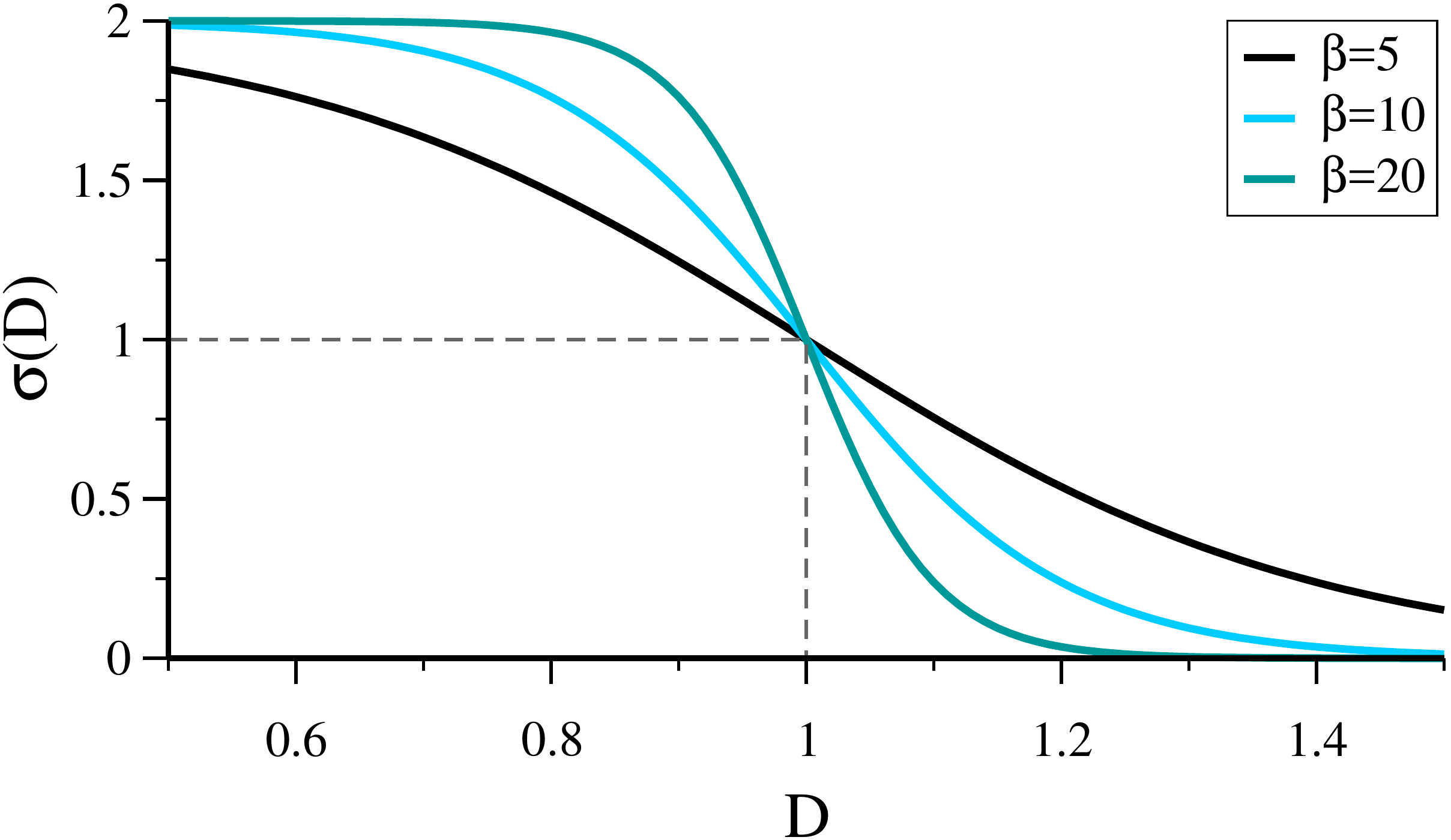}}
\caption{The rescaled Fermi function (\ref{eq:D_def})
entering the evolution
(\ref{eq:dot_V}) of the values $V$ of the electorate.
The monotonic decline of $\sigma(D)$ implies that
the desire to further increase the level $D$ of 
democratic participation drops with its actual level.
The slope at the inflection point $\sigma(1)=1$ is $-\beta/2$, 
viz proportional to the sensibility parameter $\beta$. The
time scale for opinion dynamics is hence of the order of
$2T_V/\beta$. Alternatively one may interpret the slope
and hence $\beta$ as a proxy for the rigor of 
political correctness.
}
\label{fig:fermiFunction}
\end{figure}

For the time evolution of the value $V$ we propose
\begin{equation}
T_V \frac{d}{dt} V(t) = \sigma(D(t)) - V(t),
\label{eq:dot_V}
\end{equation}
which describes a competition between a trend towards
democracy $\sim\sigma(D(t))$ and an intrinsic decay term of
the democratic values $\sim(-V(t))$. It has been observed 
in this regard that support for democratic values declines 
steadily in western societies \cite{foa2016democratic}.
If asked, to give an example, whether it is essential to live 
in a country that is governed democratically, over 70\% of US-citizens 
born around 1930 would respond yes, but only about 30\% of those 
born 1980 or later \cite{foa2016democratic}. This downward
trend translates in (\ref{eq:dot_V}) to a decay time
$T_V\approx 15-20$ years.

The actual shape of the function $\sigma(D)$ entering
(\ref{eq:dot_V}) is not relevant for the following
arguments, as long as it is monotonically declining
and hence reflecting that the desire to further increases
the current amount $D(t)$ of democratic participation
declines with its actual level. A monotonically declining 
$\sigma(D)$ incorporates therefore the notion of diminishing 
returns, which can be traced back in turn to the logarithmic 
discounting performed by the neural circuitry of the brain
\cite{dehaene2003neural,gros2012neuropsychological}.
We have chosen here for simplicity a rescaled Fermi function, 
\begin{equation}
\sigma(D) = \frac{2}{1+\exp(\beta(D-1))}~,
\label{eq:D_def}
\end{equation}
in physics jargon, for $\sigma(D)$, as illustrated in 
Fig.~\ref{fig:fermiFunction}. At the inflection point $D=1$
we have $\sigma(D=1)=1$. The parameter $\beta$, which would 
correspond to the inverse temperature in physics,
is a sensibility parameter, setting the slope
$d\sigma/dD=-\beta/2$ at the inflection point $D=1$.

The evolution equations for $D(t)$ and $V(t)$,
Eqs.~(\ref{eq:dot_D}) and (\ref{eq:dot_V}), have 
been defined such that the common fixed point
$(D,V)=(1,1)$ remains unchanged for all parameter 
settings.  This implies, that (\ref{eq:dot_D}) and (\ref{eq:dot_V})
describe the time evolution of quantities which are
relative and not bare measures. The steady-state fixed point
would evolve on the other side if $D$ and $V$ had been
measured in absolute terms \cite{spaiser2014dynamics} and 
not, as done here, relatively. The renormalization of
the steady state to $(1,1)$ does hence encompass the 
secular backdrop of declining democratic values 
\cite{foa2016democratic}. 

\begin{figure*}[t]
\centering
%
\resizebox{0.9\textwidth}{!}{\includegraphics{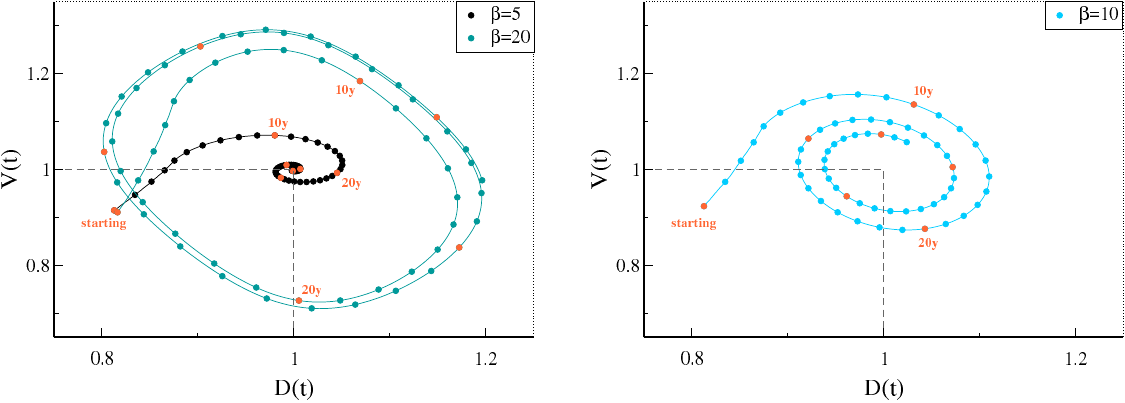}}
\caption{The result of numerically simulating (\ref{eq:dot_D}) 
and (\ref{eq:dot_V}) for $T=4$, $T_D=4$ and $T_V=15$ 
(years). The system starts (as denoted by the label `starting') 
right after the initial function, defined for $t\in[-T,0]$, 
ends, with every filled point denoting one year 
(decades are red). Note, that trajectories may intersects 
themselves for dynamical systems with time delays, as it 
happens for $\beta=20$. The fixed point $(D,V)=(1,1)$ is
stable for $\beta<\beta_c\approx 11.36$.
}
\label{fig:beta_5_10_20}
\end{figure*}

\section{Simulations results}

For the parameters entering the evolution equations
for the state of the democracy and for the values of 
the electorate, (\ref{eq:dot_D}) and (\ref{eq:dot_V})
respectively, we take $T_D=4$ years for the typical
adaption time of political actors and $T_V=15$ years for
the decay time of political values \cite{foa2016democratic}.
We start with an overview of the properties of our
model, (\ref{eq:dot_D}) together with (\ref{eq:dot_V}),
for which we set the time delay to $T=4$ years. Alternative 
values for $T$ will be considered subsequently together
with distinct ways to incorporate multiple time delays. 
For the numerical simulations we
discretized the evolution equations (\ref{eq:dot_D}) and 
(\ref{eq:dot_V}), taking one month ($\Delta t=1/12$ years) 
as a basic time step. The such obtained results do not depend
qualitatively on the exact value of $\Delta t$.

The solution of a time-delayed systems is generically contingent 
on the choice of the initial function $(D(t),V(t))$, where $t\in[-T,0]$
\cite{gros2015complex,richard2003time}. We find, however, that the system 
(\ref{eq:dot_D}) and (\ref{eq:dot_V}) is robust in the sense that
the long-time state convergences in all cases to the identical attracting 
set, which may be either a fixed point or a limit cycle, even when
fully random initial functions are selected.

In Fig.~\ref{fig:beta_5_10_20} we present typical trajectories
for $\beta=5,10,20$, where the starting function was 
$(D(t),V(t))=(0.8,0.9)$, with $t\in[-T:0]$, together with a
random jitter $\Delta D=\Delta V=0.02$. The system is stable, 
as expected, for small values of $\beta$, with the state 
$(D(t),V(t))$ of the system spiraling toward the fixed point 
$(1,1)$. The overall time-scale for the evolution is about
two decades, as consequence of $T_V=15$ year.

For an advanced democracy, characterized by a high sensibility 
$\beta=20$ to deviations from the political standard, the overall
attracting set is a limit cycle with a period of about 24.5 years
and an average deviation 
\begin{equation}
D_F = \left\langle \sqrt{\big(D(t)-1\big)^2+
                         \big(V(t)-1\big)^2}\right\rangle \approx 0.24
\label{eq:D_F}
\end{equation}
from the fixed point $(1,1)$, with the brackets $\langle\dots\rangle$
denoting the time average. In order to decide whether the limit
cycle is far away from the original fixed point, or close, we may
compare above value for $D_F$ with the functional dependency
of the response function $\sigma(D)$ entering (\ref{eq:dot_V}), 
as illustrated in Fig.~\ref{fig:fermiFunction}. We observe, 
that $D=0.8$ or $D=1.2$ leads to responses $\sigma(D)$ which 
are exponentially close to 1 and 0 respectively. This implies, 
that the limit
cycle observed for $\beta=20$ in Fig.~\ref{fig:beta_5_10_20}
is close to the maximal possible periodic solution supported
by (\ref{eq:dot_D}) and (\ref{eq:dot_V}). Even for a very
large $\beta=80$, to give an example, we find only a slightly 
increased $D_F=0.27$.

Also shown in Fig.~\ref{fig:beta_5_10_20} is a trajectory for
$\beta=10$, which spirals in the end into the fixed point 
$(D,V)=(1,1)$.  
The extraordinary long time scale needed to reach the
equilibrium state, for $\beta=10$, is a consequence of
the critical slowing down close to a phase transition,
which occurs here at $\beta_c\approx11.36$ (see discussion below).
It may hence be difficult to distinguish real-world political 
systems which are subcritical, but close to an instability,
from systems which are already unstable. 

\begin{figure*}[t]
\centering
%
\resizebox{0.9\textwidth}{!}{\includegraphics{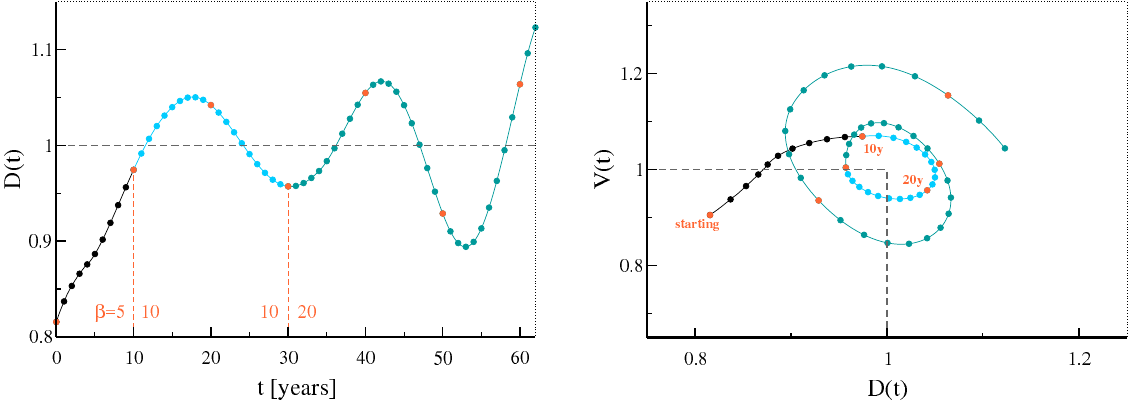}}
\caption{The result of a numerical experiment, for $T=4$, $T_D=4$ 
and $T_V=15$ (years), where $\beta=5$ for the first 10 years, 
$\beta=10$ for $t\in[10,30]$ and $\beta=20$ thereafter.
The evolution is shown for $D(t)$ as a function of time
({\it left}) and for $(D,V)$ in state space ({\it right}).
While still subcritical for $\beta=10$, the relaxation
process slows down dramatically due to the closeness 
to the phase transition occurring at $\beta_c\approx11.36$,
compare Fig.~\ref{fig:beta_5_10_20}.
}
\label{fig:beta_time}
\end{figure*}

Our basic presumption is here, that advances in communication
and organizational structures lead to a progressing optimization
of our societies which is inevitably accompanied with a decreasing
tolerance of non-standard behaviors and hence with an increasing
$\beta$, as entering (\ref{eq:dot_V}). In Fig.~\ref{fig:beta_time}
we present a scenario simulation for a time-varying $\beta$,
which is held constant at $\beta=5$ for the first ten years,
at $\beta=10$ for the subsequent twenty years and at $\beta=20$ 
thereafter. The system tries initially to reach the equilibrium
state $(D,V)=(1,1)$, being subcritical for the first thirty years,
with the relaxation towards the fixed point slowing down dramatically 
when $\beta\to10$ (compare Fig.~\ref{fig:beta_5_10_20}). Twenty years
at $\beta=10$ are not enough to equilibrate and the final
increase to $\beta=20$ leads therefore straightaway to limit-cycle 
oscillations.

\subsection{Diverging recovery times close to the Hopf bifurcation}

Normal forms allow to classify the type of bifurcations
occurring in normal dynamical systems, viz in
dynamical systems without time delays \cite{gros2015complex}.
The transition observed here at $\beta_c\approx11.36$ 
is in this context akin to a classical supercritical Hopf 
bifurcation, involving a bifurcation from a stable node 
(fixed point) to a continuously expanding periodic
orbit (stable limit cycle) \cite{piotrowska2011nature}.

In order to corroborate this statement we have evaluated
the time dependent distance $D_F(t)$ of the trajectory from 
the fixed point, as well as its long time average
(\ref{eq:D_F}). It is evident from Fig.~\ref{fig:beta_D_F},
that the size of the final limit cycle shrinks continuously
when $\beta$ approaches $\beta_c$ from above, as expected
for a second-order transition.

It is of interest to examine, for subcritical $\beta<\beta_c$,
the time scale $T_\lambda$ needed to close in to the
equilibrium state $(D,V)=(1,1)$, which is given by the 
inverse of the largest Lyapunov exponent of the
fixed point \cite{wernecke2017test}. In 
Fig.~\ref{fig:beta_D_F} we present alternatively the results 
of a numerical experiment simulating the recovery from
an external shock. For a single trajectory, with starting 
conditions as for Fig.~\ref{fig:beta_5_10_20}, the
displacement $D_F(t)$ from the steady state has been
evaluated and fitted by $\exp(-t/T_\lambda$). We notice that 
the time needed to recover from the initial displacement becomes 
of the order of three decades already for $\beta\approx 7.5$, which 
is still substantially below the critical $\beta_c\approx11.36$.
The system is hence very slow to recover from external events
pushing it away from the fixed point.

\subsection{Mixture of time delays}

\begin{figure*}[t]
\centering
%
\resizebox{0.9\textwidth}{!}{\includegraphics{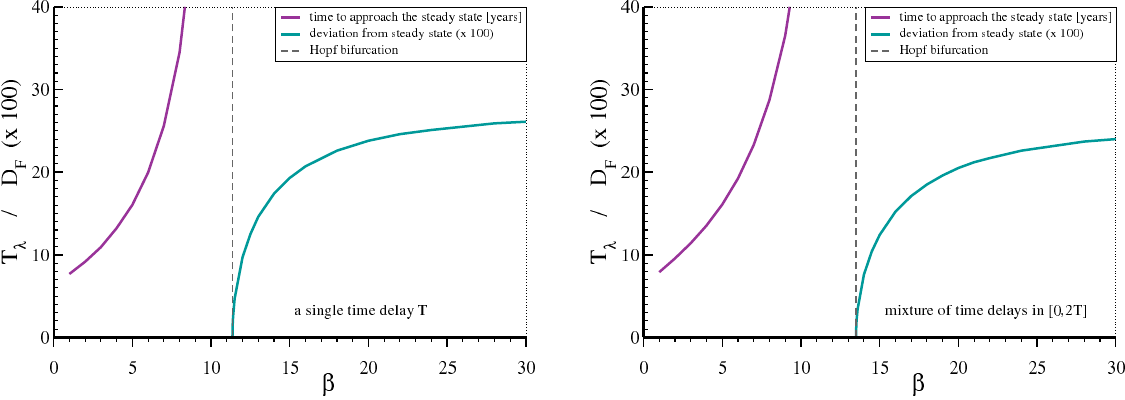}}
\caption{The results of evaluating the Euclidean distance
$D_F$ from the fixed point $(D,V)=(1,1)$. For $\beta>\beta_c$ 
(dashed vertical line) the time-average $D_F$, 
Eq.~(\ref{eq:D_F}), of the limit cycle is shown (multiplied 
by 100).
For $\beta<\beta_c$ the relaxation time $T_\lambda$ is
shown (in years). $T_\lambda$, which is also the time 
needed to recover from external shocks, has been obtained by
fitting the time-dependent Euclidean distance $D_F=D_F(t)$ by 
$\exp(-t/T_\lambda)$. The data is for the model with
a single time delay $T=4$ (Eq.~(\ref{eq:dot_D}), {\it left panel})
and for the model with a uniform mixture of time delays
(Eq.~(\ref{eq:dot_D_mixture}), {\it right panel}) and
otherwise identical parameters. The respective critical
sensitivities are $\beta_c\approx11.36$ ({\it left}) and
$\beta_c\approx13.5$ ({\it right}).
}   
\label{fig:beta_D_F}
\end{figure*}

With (\ref{eq:dot_D}) we assumed that the state 
$D(t)$ tries to align itself to values the
electorate expressed exactly $T$ years before. 
A mixture of time delays may contribute in reality. 
We consider with 
\begin{equation}
T_D \frac{d}{dt} D(t) \ =\   \overline V_\alpha(t) - D(t)
\label{eq:dot_D_mixture}
\end{equation}
the coupling of $D(t)$ to two specific distributions
$\alpha=1,2$ of lag times,
\begin{eqnarray}
\label{eq:dot_D_mixture_1}
\overline V_1(t) &=& \frac{1}{2T}\int_0^{2T} V(t-\tau)d\tau \\
\overline V_2(t) &=& \frac{1}{T}\int_0^{\infty} \mathrm{e}^{-\tau/T}V(t-\tau)d\tau
\label{eq:dot_D_mixture_2}
\end{eqnarray}
where $V_1(t)$ and $V_2(t)$ correspond respectively to a flat 
distribution, with $T\in [0,2T]$, and to exponentially discounted 
delay times. The average time delay stays at $T$ in both cases. 
We find, as shown in Fig.~\ref{fig:beta_D_F}, that a flat distribution,
viz $\overline V_1$ in Eq.~(\ref{eq:dot_D_mixture}),
induces only relative minor quantitative changes, with
all qualitative features of the original model (\ref{eq:dot_D}) 
remaining untouched. There is a slight upward renormalization, 
when using $\overline V_1$, of the critical sensitivity from 
$\beta_c\approx11.36$, as obtained for
(\ref{eq:dot_D}), to $\beta_c\approx13.5$.

For exponentially discounted lag times, describing the common but
not exclusive case that past messages are progressively discounted 
in the context of political communication \cite{chong2010dynamic}, 
we find numerically that $\beta_c\approx 23.7$, which is now substantially 
increased, but otherwise no overall qualitative changes. 

\section{Stability analysis}

\begin{figure*}[t]
\centering
%
\resizebox{0.9\textwidth}{!}{\includegraphics{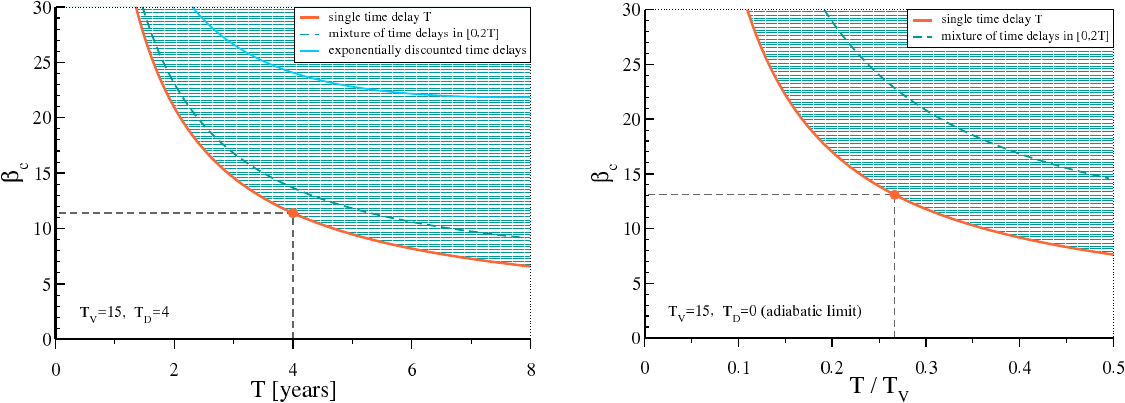}}
\caption{The Hopf bifurcation line for the case of a
single time delay (full red curve), for  a uniform distribution 
of delay times (dashed cyan curve) and for exponentially
distributed time delays (full blue curve).
The attracting state is a limit cycle above the respective
lines (viz in the shaded region for the case of a single
time delay), and a fixed point otherwise. The dashed rectangle
indicates the case of a single time delay $T=4$, and an
adaption time scale for $V(t)$ of $T_V=15$. 
{\it Left:} For $T_D=4$ and the original model (\ref{eq:dot_D}) 
and (\ref{eq:dot_V}), respectively (\ref{eq:dot_D_mixture}),
for which the bifurcations lines $\beta_c$ are determined by 
(\ref{eq:Hopf_full_beta_c}), (\ref{eq:Hopf_mix_beta_c}) and 
(\ref{eq:beta_c_exp_discounted}).
{\it Right:} For the adiabatic limit (\ref{eq:dot_V_adiabatic}), 
obtained when $T_D\to0$. In this limit there is no
Hopf bifurcation for exponentially discounted time delays.
}   
\label{fig:PD_all}
\end{figure*}

The stability of the fixed point $(D,V)=(1,1)$ can be examined
\cite{boukas2012deterministic} by linearizing the evolution
equations (\ref{eq:dot_D}) and (\ref{eq:dot_V}) 
\begin{eqnarray}
\label{eq:D_dot_linearized}
T_D\frac{d}{dt}\delta D(t) &=& \delta V(t-T)-\delta D(t),\\
T_V\frac{d}{dt}\delta V(t) &=& -\frac{\beta}{2}\delta D(t)-\delta V(t)~,
\label{eq:V_dot_linearized}
\end{eqnarray}
where $\delta D= D-1$ and $\delta V= V-1$. The Ansatz
$\delta D(t) = D_0\exp(\lambda t)$ and
$\delta V(t) = V_0\exp(\lambda t)$ leads to 
$$
V_0 \mathrm{e}^{-\lambda T} = D_0(1+T_D\lambda),
\qquad
D_0 = -\frac{2V_0}{\beta}(1+T_V\lambda)~,
$$
and hence to
\begin{equation}
\mathrm{e}^{-\lambda T} = -\frac{2}{\beta}(1+T_V\lambda)(1+T_D\lambda)~.
\label{eq:DV_lambda}
\end{equation}
The Lyapunov exponent $\lambda=\lambda'+i\lambda''$ is
generically complex, becoming purely imaginary, with
$\lambda'=0$, at the bifurcation $\beta\to\beta_c$.
The real and imaginary components of (\ref{eq:DV_lambda})
then are:
\begin{eqnarray}
\label{eq:DV_real}
\cos(\lambda''T) &=& -\frac{2}{\beta_c}\left(1-T_VT_D(\lambda'')^2\right),
\\
\sin(\lambda''T) &=& \frac{2}{\beta_c}(T_V+T_D)\lambda''~,
\label{eq:DV_imag}
\end{eqnarray}
or
\begin{equation}
\tan(T\lambda'') = \frac{(T_D+T_V)\lambda''}{T_DT_V(\lambda'')^2-1},
\label{eq:Hopf_full_lambda}
\end{equation}
and
\begin{equation}
\frac{\beta_c^2}{4} 
= \left(1+(T_D\lambda'')^2\right)\left(1+(T_V\lambda'')^2\right)~,
\label{eq:Hopf_full_beta_c}
\end{equation}
where we have used that 
\begin{eqnarray*}
(T_DT_V(\lambda'')^2-1)^2 &+& (T_D+T_V)^2 (\lambda'')^2 \\
&=& (1+(T_D\lambda'')^2)
(1+(T_V\lambda'')^2)~.
\end{eqnarray*}
One solves first (\ref{eq:Hopf_full_lambda}) for 
$\lambda''$ and then (\ref{eq:Hopf_full_beta_c}) for $\beta_c$. 
The corresponding 
phase diagram is presented in Fig.~\ref{fig:PD_all} for fixed 
$T_D=4$ and $T_V=15$. The locus of the phase transition at $T=4$ 
is $\beta_c=11.4$, which  differs only marginally from the one 
found in the numerical simulation, $\beta_c=11.36$, for 
which time had been discretized (using $\Delta t=1/12$).

\subsection{Uniform mixture of time delays}

For the case (\ref{eq:dot_D_mixture}) of an uniform
mixture of time delays one replaces
$\exp(-\lambda T)$ in (\ref{eq:DV_lambda}) by
$\int \exp(-\lambda \tau)d\tau/(2T)$, obtaining
\begin{eqnarray*}
\frac{1}{2T}\int_0^{2T} \cos(\lambda''\tau) d\tau
&=& \frac{2}{\beta_c}\big[T_VT_D(\lambda'')^2-1\big],
\\
\frac{1}{2T}\int_0^{2T} \sin(\lambda''\tau) d\tau
&=& \frac{2}{\beta_c}(T_V+T_D)\lambda''~,
\end{eqnarray*}
which results in turn, after carrying out the 
respective integrals, in 
\begin{eqnarray*}
\sin(2T\lambda'')
&=& \frac{4T\lambda''}{\beta_c}\big[T_VT_D(\lambda'')^2-1\big],
\\
1-\cos(2T\lambda'')
&=& \frac{4T\lambda''}{\beta_c}(T_V+T_D)\lambda''~.
\end{eqnarray*}
With $\sin(2\lambda'')=2\sin(T\lambda'')\cos(T\lambda'')$ and
$\cos(2T\lambda'')=1-2\sin^2(T\lambda'')$ we then obtain
\begin{equation}
\tan(T\lambda'') = \frac{(T_D+T_V)\lambda''}{T_DT_V(\lambda'')^2-1}
\label{eq:Hopf_mix_lambda}
\end{equation}
and
\begin{equation}
\frac{\beta_c}{2}\frac{T_D+T_V}{T} =
\left(1+(T_D\lambda'')^2\right)\left(1+(T_V\lambda'')^2\right)~.
\label{eq:Hopf_mix_beta_c}
\end{equation}
Note that the expressions (\ref{eq:Hopf_mix_lambda})
and (\ref{eq:Hopf_full_lambda}) for the imaginary 
component $\lambda''$ of the Lyapunov exponents
are identical and, correspondingly, also the
right-hand sides of (\ref{eq:Hopf_mix_beta_c})
and (\ref{eq:Hopf_full_beta_c}).
For above transformations we used 
$$
[1-\cos(2T\lambda'')]/\sin(2T\lambda'')=\tan(T\lambda'')
$$ 
and that
\begin{eqnarray*}
4\sin^2(T\lambda'') &=&
\frac{(4T\lambda'')^2}{\beta_c^2} 
\left(1+(T_D\lambda'')^2\right)\left(1+(T_V\lambda'')^2\right)
\\ &= &
\frac{4\tan^2(T\lambda'')}{1+\tan^2(T\lambda'')}
\end{eqnarray*}
can be simplified when using (\ref{eq:Hopf_mix_lambda}).
The bifurcation line resulting from (\ref{eq:Hopf_mix_beta_c}), which has 
been included in Fig.~\ref{fig:PD_all}, runs somewhat parallel  to 
the one obtained via (\ref{eq:Hopf_full_beta_c}) for the case of a single 
delay time, closing in for $T\ll T_V$, when the actual distribution
of lag times becomes unimportant. For $T=4$ we find that $\beta_c$ 
increases from $\beta_c=11.4$ to $\beta_c=13.68$.

Comparing (\ref{eq:Hopf_full_lambda}) and (\ref{eq:Hopf_mix_lambda}) one
finds, remarkably, that the imaginary part $\lambda''$ of 
the Lyapunov exponent is identical at criticality, albeit
at different values of $\beta_c$. This implies, that the
revolution frequencies of the resulting limit cycles are
identical in the respective limits $\beta\to\beta_c$ 
from above.

\subsection{Exponentially discounted time delays}

For exponentially discounted delay times (\ref{eq:dot_D_mixture}) 
we need
\begin{eqnarray*}
\frac{1}{T}\int_0^{2T} \mathrm{e}^{-t/T}\cos(\lambda''\tau) d\tau
&=& \frac{1}{1+(T\lambda'')^2},
\\
\frac{1}{T}\int_0^{2T} \mathrm{e}^{-t/T}\sin(\lambda''\tau) d\tau
&=& \frac{T\lambda''}{1+(T\lambda'')^2}~,
\end{eqnarray*}
which results respectively in
\begin{eqnarray}
\label{eq:DV_exp_discounted_real}
\frac{1}{1+(T\lambda'')^2} &=& \frac{2}{\beta_c}\left(T_VT_D(\lambda'')^2-1\right),
\\
\frac{1}{1+(T\lambda'')^2} &=& \frac{2}{\beta_c}\frac{T_V+T_D}{T}
\label{eq:DV_exp_discounted_imag}
\end{eqnarray}

\noindent
instead of (\ref{eq:DV_real}) and (\ref{eq:DV_imag}). We then find
\begin{eqnarray}
\nonumber
(\lambda'')^2&=&\frac{T_D+T_V+T}{T_DT_VT} \\
\beta_c &=& 2 \frac{T_D+T_V}{T}\big[1+(T\lambda'')^2\big]
\label{eq:beta_c_exp_discounted}
\end{eqnarray}
for the Hopf bifurcation line. The critical $\beta_c$ has been 
included in Fig.~\ref{fig:PD_all}. For $T_D=4=T$ and $T_V=15$ the 
resulting $\beta_c=24.1$ is again marginally larger than 
the value, $\beta_c\approx23.7$, obtained from corresponding 
time discretized numerical simulation.

\subsection{Adiabatic limit} 

We have shown above that our model is robust against
changes in the distribution of time delays. The nature
of the attracting states are also not sensitively 
dependent on the ratio of $T_D/T_V$. It is illustrative, 
in this context, to examine the adiabatic limit $T_D\ll T_V$ 
of (\ref{eq:dot_D}) and (\ref{eq:dot_V}), for which 
$D(t)$ follows closely $V(t-T)$. In this case one
can substitute $D(t)$ by $V(t-T)$ in (\ref{eq:dot_V}),
obtaining
\begin{equation}
T_V \frac{d}{dt} V(t) \ =\ \sigma(V(t-T)) - V(t)~.
\label{eq:dot_V_adiabatic}
\end{equation}
The locus of the bifurcation is determined by 
(\ref{eq:Hopf_full_beta_c}) in the limit $T_D\to0$, 
or, alternatively, by
\begin{equation}
\tan(x\,T/T_V) = -x,
\qquad
\frac{\beta_c^2}{4}=1+x^2,
\qquad
x = T_V\lambda'' ~,
\label{eq:beta_c_adiabatic}
\end{equation}
when using rescaled variables. $\beta_c$ is then dependent only
on the ratio $T/T_V$, as shown in Fig.~\ref{fig:PD_all}. For the
case of a uniform mixture of time delays (\ref{eq:Hopf_mix_lambda})
and (\ref{eq:Hopf_mix_beta_c}) reduce to 
\begin{equation}
\tan(x\,T/T_V) = -x,
\qquad\quad
\frac{\beta_c}{2}\frac{T_V}{T}=1+x^2
\label{eq:beta_c_adiabatic_mix}
\end{equation}
in the limit $T_D\to0$. One notices, compare Fig.~\ref{fig:PD_all},
that there is a substantial quantitative difference in the
adiabatic limit between having a single and a mixture of
time delays. 

Interestingly, there is no phase transition in the adiabatic limit 
for the case of exponentially discounted time delays,
with (\ref{eq:DV_exp_discounted_real}) having no solution
in the limit $T_D\to0$.

\subsection{Properties of the phase diagram}

The phase diagrams presented in Fig.~\ref{fig:PD_all}
have a series of common features.

\begin{itemize}
\item The Hopf bifurcation line is a monotonically decreasing 
      function. For small
      time delays $T$ one needs a higher sensibility
      $\beta>\beta_c$ for the instability to occur,
      and vice verse.
\item There is no minimal time delay $T$, viz there
      is a critical $\beta_c<\infty$ for any $T>0$,
      with
\begin{equation}
\lim_{T/T_V\to0} \beta_c(T/T_V) \to \infty~.
\label{eq:beta_c_limit}
\end{equation}
     The fixed point is hence stable for all $\beta$ when
     there is no time delay, $T=0$.
\item There is a lower $\beta_c$ below which the fixed point
      is stable even when $T$ is arbitrary large. In the
      adiabatic limit (\ref{eq:beta_c_adiabatic}) one needs
      $\beta_c>2$. 
\item The imaginary part $\lambda''$ of the Lyapunov exponent
      needs to be non-zero for (\ref{eq:Hopf_full_lambda}) and 
      (\ref{eq:Hopf_mix_lambda}) to have a non-trivial solution.
      $\lambda''$ is hence finite at the transition, the 
      tell-sign of a Hopf bifurcation \cite{gros2015complex}.
      The revolution frequency of the limit cycle, which
      is of the order of $1/|\lambda''|$, is hence not
      critical, varying smoothly above the transition.
\end{itemize}
In the vicinity of the transition the sensibility $\beta$ 
induces a speed-up of the reactive value dynamics, as evident 
from the linearized equations (\ref{eq:D_dot_linearized}) and 
(\ref{eq:V_dot_linearized}), by a factor $\beta/2$, which may 
be identified with a corresponding acceleration of opinion 
dynamics. The overall time needed to reach the fixed point 
nevertheless diverges as $1/\lambda'\sim 1/|\beta-\beta_c|$.
This phenomenon, known as critical slowing down, is 
observed generically in dynamical systems close to a tipping 
point. It is observed in a wide range of settings, affecting, 
e.g., the resilience of ecosystems \cite{van2007slow} as well 
as the evolution of the climate prior to a major shift 
\cite{dakos2008slowing}. The increased time scales needed 
to react to disturbances close the instability are also 
evident in Fig.~\ref{fig:beta_D_F}.


\section{Discussion}

There are two mutually not exclusive routes 
to describe the conflict between slow political
decision making and accelerating social dynamics 
\cite{goetz2014question,tomba2014clash,rosa2013social}. 
In the first view politics continuously adapts, over the 
course of $T_D$ years, to the current demands of the 
electorate. Time lags are absent in this scenario
and the system stable for all parameters. Politics
then evolves around a stable state, with deviations 
from the fixed point driven exclusively by external events.

Here we have examined a second possibility, namely that 
a certain fraction of political decision making results from
the response to demands the electorate voiced $T$ years 
ago. The time delay $T$ may be either fixed or drawn 
from a continuous distribution, as described by 
Eqs.~(\ref{eq:dot_D}) and (\ref{eq:dot_D_mixture}) respectively. 
For both cases we find that the socio-political system becomes 
inherently unstable whenever the electorate responds 
sensitively to political changes. This conclusion, which is 
robust and independent of the details of the here used model,
results from the fact that time delays will inherently amplify 
fluctuations once their influence becomes substantial.

In our model the sensitivity $\beta$ of the electorate
leads to typical reaction times $2T_V/\beta$, as evident 
form the linearized evolution equation (\ref{eq:V_dot_linearized}),
where $T_V$ is the time scale for the long-term evolution
of basic political values. In order to obtain estimates for
real-world political communication we considered the
case of exponentially discounted time delays, for which
the instability occurs at 
$\beta_c\approx24.1$ for $T=4$ and at
$\beta_c\approx50.7$ for $T=1$ (compare Fig.~\ref{fig:PD_all}). 
Socio-political instabilities then start to manifest 
themselves for $T=4$ when the corresponding times 
scale $2T_V/\beta_c$ for the opinion dynamics falls 
below $30/24.1$ years (about 15 months). For a time
delay of one year, $T=1$, instabilities develop when
the opinion dynamics takes place on time scale below 
$30/50.7$ years (about 7 months). 

Our estimates for the tipping point of political opinion 
dynamics, 7-15 months when assuming mean time delays of 
the order of 1-4 years, are for aggregate processes which 
include the effects of fast news propagation as well as
the consequences of slowly but continuously changing 
preset political beliefs. It is conceivable within out
model that western democracies have seen the unfolding of
a slow but steady long-term acceleration of opinion dynamics,
with the passing of the threshold of 7-15 months contributing 
to the recent emergence of political styles disrupting 
political conventions considered hitherto as fundamental 
\cite{inglehart2016trump}.
External effects, such as the 2007-08 financial crisis 
\cite{shiller2012subprime,funke2016going}, would induce
in this view an additional temporary but sharp rise in $\beta$.

An important aspect regards the time needed to recover from
an external disrupting event, such as a global crisis. 
Naively one may expect that the accelerating pace of opinion 
formation observed in advanced democracies would reduce 
typical recovery times. The contrary is however the case.
It is well known, as illustrated in Fig.~\ref{fig:beta_D_F},
that second order instabilities lead to critical slowing down 
in their proximity and hence to diverging recovery times.  
As a consequence one observes long-lasting oscillations even
below the actual transition, illustrated shown in Fig.~\ref{fig:beta_5_10_20}.
Analogous oscillations matching both the period (about 
20 years), and the magnitude (10\%-15\%), have be observed
since the early 1990th in Australian polls studying aggregate 
value orientations along the materialism vs.\ postmaterialism axis
\cite{tranter2015impact}. A substantially larger corpus of
data would however been needed for an eventual validation, or
falsification, of the here presented approach. Note that
our framework describes instabilities arising within 
representative democracies and not transitions to 
non-democratic regimes.

The scope of the work presented here is to point
out a phenomenon of possible key importance for
the understanding of the long-term stability of 
representative democracies. The instabilities we 
find lead to oscillatory but not to irregular 
socio-political states. One possibility to extend 
our study would however be to consider time delays 
varying periodically with the election cycle. It
is to be expected that such kinds of non-constant
time delays would act as periodic drivings 
\cite{d1982chaotic}, which are in turn known to 
induce transitions to chaotic states in non-linear 
dynamical systems. We note in this context that 
transitions to potentially disrupting states with 
runaway opinion growth have been observed \cite{podobnik2017predicting} 
in agent based simulations examining the response of 
an electorate to rising levels of immigration.

\subsection*{Acknowledgments}

We thank Karolin Kappler regarding discussions concerning
social acceleration,  Daniel Lambach regarding time
delays in democratic structures and Roser Valenti for
reading the manuscript.

\bibliographystyle{unsrt}

\begin{thebibliography}{10}

\bibitem{erneux2009applied}
Thomas Erneux.
\newblock {\em Applied delay differential equations}, volume~3.
\newblock Springer Science \& Business Media, 2009.

\bibitem{gros2015complex}
C.~Gros.
\newblock {\em Complex and adaptive dynamical systems: A primer}.
\newblock Springer, 2015.

\bibitem{singh2012just}
Gurinder Singh and Inderpreet~Singh Ahuja.
\newblock Just-in-time manufacturing: literature review and directions.
\newblock {\em International Journal of Business Continuity and Risk
  Management}, 3(1):57--98, 2012.

\bibitem{schnellenbach2015behavioral}
Jan Schnellenbach and Christian Schubert.
\newblock Behavioral political economy: A survey.
\newblock {\em European Journal of Political Economy}, 40:395--417, 2015.

\bibitem{wolin1997time}
Sheldon~S Wolin.
\newblock What time is it?
\newblock {\em Theory \& Event}, 1(1), 1997.

\bibitem{goetz2014question}
Klaus~H Goetz.
\newblock A question of time: Responsive and responsible democratic politics.
\newblock {\em West European Politics}, 37(2):379--399, 2014.

\bibitem{fleischer2013time}
Julia Fleischer.
\newblock Time and crisis.
\newblock {\em Public Management Review}, 15(3):313--329, 2013.

\bibitem{tomba2014clash}
Massimiliano Tomba.
\newblock Clash of temporalities: Capital, democracy, and squares.
\newblock {\em South Atlantic Quarterly}, 113(2):353--366, 2014.

\bibitem{wolffsohn2001nomen}
Michael Wolffsohn and Thomas Brechenmacher.
\newblock Nomen est omen: The selection of first names as an indicator for
  public opinion in the past.
\newblock {\em International Journal of Public Opinion Research},
  13(2):116--139, 2001.

\bibitem{rosa2013social}
Hartmut Rosa.
\newblock {\em Social acceleration: A new theory of modernity}.
\newblock Columbia University Press, 2013.

\bibitem{pain2017unusual}
Elisabeth Pain.
\newblock Unusual presidential race rattles french scientists, 2017.

\bibitem{menon2016brexit}
Anand Menon and JOHN-PAUL SALTER.
\newblock Brexit: initial reflections.
\newblock {\em International Affairs}, 92(6):1297--1318, 2016.

\bibitem{dalton2004advanced}
Russell~J Dalton, Susan~E Scarrow, and Bruce~E Cain.
\newblock Advanced democracies and the new politics.
\newblock {\em Journal of democracy}, 15(1):124--138, 2004.

\bibitem{hughes2011political}
Geoffrey Hughes.
\newblock {\em Political correctness: a history of semantics and culture}.
\newblock John Wiley \& Sons, 2011.

\bibitem{maass2013does}
Anne Maass, Caterina Suitner, and Elisa Merkel.
\newblock Does political correctness make (social) sense?
\newblock In Joseph~P Forgas, Orsolya Vincze, and Janos Laszlo, editors, {\em
  Social Cognition and communication}, pages 331--346. Psychology Press New
  York, NY, 2013.

\bibitem{abdollahian2012dynamics}
Mark~A Abdollahian, Travis~G Coan, Hana Oh, and Birol~A Yesilada.
\newblock Dynamics of cultural change: the human development perspective.
\newblock {\em International Studies Quarterly}, 56(4):827--842, 2012.

\bibitem{spaiser2014dynamics}
Viktoria Spaiser, Shyam Ranganathan, Richard~P Mann, and David~JT Sumpter.
\newblock The dynamics of democracy, development and cultural values.
\newblock {\em PloS one}, 9(6):e97856, 2014.

\bibitem{alexander2012measuring}
Amy~C Alexander, Ronald Inglehart, and Christian Welzel.
\newblock Measuring effective democracy: A defense.
\newblock {\em International Political Science Review}, 33(1):41--62, 2012.

\bibitem{ranganathan2014bayesian}
Shyam Ranganathan, Viktoria Spaiser, Richard~P Mann, and David~JT Sumpter.
\newblock Bayesian dynamical systems modelling in the social sciences.
\newblock {\em PloS one}, 9(1):e86468, 2014.

\bibitem{inglehart2016trump}
Ronald Inglehart and Pippa Norris.
\newblock Trump, brexit, and the rise of populism: Economic have-nots and
  cultural backlash.
\newblock 2016.

\bibitem{foa2016democratic}
Roberto~Stefan Foa and Yascha Mounk.
\newblock The democratic disconnect.
\newblock {\em Journal of Democracy}, 27(3):5--17, 2016.

\bibitem{dehaene2003neural}
Stanislas Dehaene.
\newblock The neural basis of the weber--fechner law: a logarithmic mental
  number line.
\newblock {\em Trends in cognitive sciences}, 7(4):145--147, 2003.

\bibitem{gros2012neuropsychological}
Claudius Gros, Gregor Kaczor, and Dimtrij{\'e} Markovi{\'c}.
\newblock Neuropsychological constraints to human data production on a global
  scale.
\newblock {\em The European Physical Journal B}, 85(1):1--5, 2012.

\bibitem{richard2003time}
Jean-Pierre Richard.
\newblock Time-delay systems: an overview of some recent advances and open
  problems.
\newblock {\em automatica}, 39(10):1667--1694, 2003.

\bibitem{piotrowska2011nature}
Monika~J Piotrowska and Urszula Fory{\'s}.
\newblock The nature of hopf bifurcation for the gompertz model with delays.
\newblock {\em Mathematical and Computer Modelling}, 54(9):2183--2198, 2011.

\bibitem{wernecke2017test}
Hendrik Wernecke, Bulcs{\'u} S{\'a}ndor, and Claudius Gros.
\newblock How to test for partially predictable chaos.
\newblock {\em Scientific reports}, 7(1):1087, 2017.

\bibitem{chong2010dynamic}
Dennis Chong and James~N Druckman.
\newblock Dynamic public opinion: Communication effects over time.
\newblock {\em American Political Science Review}, 104(04):663--680, 2010.

\bibitem{boukas2012deterministic}
El-Kebir Boukas and Zi-Kuan Liu.
\newblock {\em Deterministic and stochastic time-delay systems}.
\newblock Springer Science \& Business Media, 2012.

\bibitem{van2007slow}
Egbert~H Van~Nes and Marten Scheffer.
\newblock Slow recovery from perturbations as a generic indicator of a nearby
  catastrophic shift.
\newblock {\em The American Naturalist}, 169(6):738--747, 2007.

\bibitem{dakos2008slowing}
Vasilis Dakos, Marten Scheffer, Egbert~H van Nes, Victor Brovkin, Vladimir
  Petoukhov, and Hermann Held.
\newblock Slowing down as an early warning signal for abrupt climate change.
\newblock {\em Proceedings of the National Academy of Sciences},
  105(38):14308--14312, 2008.

\bibitem{shiller2012subprime}
Robert~J Shiller.
\newblock {\em The subprime solution: how today's global financial crisis
  happened, and what to do about it}.
\newblock Princeton University Press, 2012.

\bibitem{funke2016going}
Manuel Funke, Moritz Schularick, and Christoph Trebesch.
\newblock Going to extremes: Politics after financial crises, 1870--2014.
\newblock {\em European Economic Review}, 2016.

\bibitem{tranter2015impact}
Bruce Tranter.
\newblock The impact of political context on the measurement of postmaterial
  values.
\newblock {\em Sage Open}, 5(2):2158244015591826, 2015.

\bibitem{d1982chaotic}
D~d'Humieres, MR~Beasley, BA~Huberman, and A~Libchaber.
\newblock Chaotic states and routes to chaos in the forced pendulum.
\newblock {\em Physical Review A}, 26(6):3483, 1982.

\bibitem{podobnik2017predicting}
Boris Podobnik, Marko Jusup, Dejan Kovac, and HE~Stanley.
\newblock Predicting the rise of eu right-wing populism in response to
  unbalanced immigration.
\newblock {\em Complexity}, 2017, 2017.

\end{thebibliography}

\end{document}